\documentclass[%
reprint,
amsmath,amssymb,
aps,
]{revtex4-2}

\usepackage{feynmp-auto}

\usepackage{graphicx}% Include figure files
\usepackage{dcolumn}% Align table columns on decimal point
\usepackage{bm}% bold math
\usepackage{hyperref}% add hypertext capabilities
\usepackage{color,soul}

\usepackage{
tikz,
pgfplots
}

\usepackage{lipsum}

\hypersetup{
colorlinks=true
,citecolor=red
,urlcolor=blue
,anchorcolor=blue
,citecolor=red
,filecolor=blue
,linkcolor=blue
,menucolor=blue
,linktocpage=true
,pdfproducer=medialab
}

\allowdisplaybreaks

\newcommand{\GraphTolopogyTone}{
    \parbox[c][10mm][c]{11mm}{
        \centering
        \begin{fmfgraph*}(10,10)
            \fmfleft{i}
            \fmfright{o}
            \fmf{plain}{i,v1}
            \fmf{plain}{o,v2}
            \fmffreeze
            \fmfforce{(0.1w,0.5h)}{v1}
            \fmfforce{(0.9w,0.5h)}{v2}
            \fmf{plain,left=1}{v1,v2}
            \fmf{plain,right=1}{v1,v2}
        \end{fmfgraph*}
    }
}

\newcommand{\GraphTolopogyTtwo}{
    \parbox[c][10mm][c]{11mm}{
        \centering
        \begin{fmfgraph*}(10,10)
            \fmfleft{i}
            \fmfright{o}
            \fmf{plain}{i,v1}
            \fmf{plain}{o,v2}
            \fmffreeze
            \fmfforce{(0.1w,0.5h)}{v1}
            \fmfforce{(0.9w,0.5h)}{v2}
            \fmf{plain,left=1,tag=1}{v1,v2}
            \fmf{plain,right=1,tag=2}{v1,v2}
            \fmfposition
            \fmfipath{p[]}
            \fmfiset{p1}{vpath1(__v1,__v2)}
            \fmfiset{p2}{vpath2(__v2,__v1)}
            \fmfi{plain}{point length(p1)/2 of p1 -- point length(p2)/2 of p2}
        \end{fmfgraph*}
    }
}

\newcommand{\GraphTolopogyLthree}{
    \parbox[c][10mm][c]{11mm}{
        \centering
        \begin{fmfgraph*}(10,10)
            \fmfleft{i}
            \fmfright{o}
            \fmf{plain}{i,v1}
            \fmf{plain}{o,v2}
            \fmffreeze
            \fmfforce{(0.1w,0.5h)}{v1}
            \fmfforce{(0.9w,0.5h)}{v2}
            \fmf{plain,left=1,tag=1}{v1,v2}
            \fmf{plain,right=1,tag=2}{v1,v2}
            \fmfposition
            \fmfipath{p[]}
            \fmfiset{p1}{vpath1(__v1,__v2)}
            \fmfiset{p2}{vpath2(__v2,__v1)}
            \fmfi{plain}{point 1*length(p1)/3 of p1 -- point 1*length(p2)/3 of p2}
            \fmfi{plain}{point 2*length(p1)/3 of p1 -- point 2*length(p2)/3 of p2}
        \end{fmfgraph*}
    }
}

\newcommand{\GraphTolopogyBthree}{
    \parbox[c][10mm][c]{11mm}{
        \centering
        \begin{fmfgraph*}(10,10)
            \fmfleft{i}
            \fmfright{o}
            \fmf{plain}{i,v1}
            \fmf{plain}{o,v2}
            \fmffreeze
            \fmfforce{(0.1w,0.5h)}{v1}
            \fmfforce{(0.9w,0.5h)}{v2}
            \fmf{plain,left=1,tag=1}{v1,v2}
            \fmf{plain,right=1,tag=2}{v1,v2}
            \fmf{phantom,tag=3}{v1,v2}
            \fmfposition
            \fmfipath{p[]}
            \fmfiset{p1}{vpath1(__v1,__v2)}
            \fmfiset{p2}{vpath2(__v2,__v1)}
            \fmfiset{p3}{vpath3(__v1,__v2)}
            \fmfi{plain}{point 1*length(p1)/4 of p1 -- point 1*length(p3)/2 of p3}
            \fmfi{plain}{point 3*length(p1)/4 of p1 -- point 1*length(p3)/2 of p3}
            \fmfi{plain}{point 1*length(p3)/2 of p3 -- point 1*length(p2)/2 of p2}
        \end{fmfgraph*}
    }
}

\newcommand{\GraphTolopogyNthree}{
    \parbox[c][10mm][c]{11mm}{
        \centering
        \begin{fmfgraph*}(10,10)
            \fmfleft{i}
            \fmfright{o}
            \fmf{plain}{i,v1}
            \fmf{plain}{o,v2}
            \fmffreeze
            \fmfforce{(0.1w,0.5h)}{v1}
            \fmfforce{(0.9w,0.5h)}{v2}
            \fmf{plain,left=1,tag=1}{v1,v2}
            \fmf{plain,right=1,tag=2}{v1,v2}
            \fmfposition
            \fmfipath{p[]}
            \fmfiset{p1}{vpath1(__v1,__v2)}
            \fmfiset{p2}{vpath2(__v2,__v1)}
            \fmfi{plain,rubout=3}{point 1*length(p1)/4 of p1 -- point 3*length(p2)/4 of p2}
            \fmfi{plain,rubout=3}{point 3*length(p1)/4 of p1 -- point 1*length(p2)/4 of p2}
        \end{fmfgraph*}
    }
}

\newcommand{\GraphTolopogyLfour}{
    \parbox[c][10mm][c]{11mm}{
        \centering
        \begin{fmfgraph*}(10,10)
            \fmfleft{i}
            \fmfright{o}
            \fmf{plain}{i,v1}
            \fmf{plain}{o,v2}
            \fmffreeze
            \fmfforce{(0.1w,0.5h)}{v1}
            \fmfforce{(0.9w,0.5h)}{v2}
            \fmf{plain,left=1,tag=1}{v1,v2}
            \fmf{plain,right=1,tag=2}{v1,v2}
            \fmfposition
            \fmfipath{p[]}
            \fmfiset{p1}{vpath1(__v1,__v2)}
            \fmfiset{p2}{vpath2(__v2,__v1)}
            \fmfi{plain}{point 1*length(p1)/3 of p1 -- point 1*length(p2)/3 of p2}
            \fmfi{plain}{point 1*length(p1)/2 of p1 -- point 1*length(p2)/2 of p2}
            \fmfi{plain}{point 2*length(p1)/3 of p1 -- point 2*length(p2)/3 of p2}
        \end{fmfgraph*}
    }
}

\newcommand{\GraphTolopogyHfour}{
    \parbox[c][10mm][c]{11mm}{
        \centering
        \begin{fmfgraph*}(10,10)
            \fmfleft{i}
            \fmfright{o}
            \fmf{plain}{i,v1}
            \fmf{plain}{o,v2}
            \fmffreeze
            \fmfforce{(0.1w,0.5h)}{v1}
            \fmfforce{(0.9w,0.5h)}{v2}
            \fmf{plain,left=1,tag=1}{v1,v2}
            \fmf{plain,right=1,tag=2}{v1,v2}
            \fmf{phantom,tag=3}{v1,v2}
            \fmfposition
            \fmfipath{p[]}
            \fmfiset{p1}{vpath1(__v1,__v2)}
            \fmfiset{p2}{vpath2(__v2,__v1)}
            \fmfiset{p3}{vpath3(__v2,__v1)}
            \fmfi{plain}{point 1*length(p3)/4 of p3 .. point 3*length(p3)/4 of p3}
            \fmfi{plain}{point 1*length(p1)/3 of p1 .. point 1*length(p2)/3 of p2}
            \fmfi{plain}{point 2*length(p1)/3 of p1 .. point 2*length(p2)/3 of p2}
        \end{fmfgraph*}
    }
}

\newcommand{\GraphTolopogyFfour}{
    \parbox[c][10mm][c]{11mm}{
        \centering
        \begin{fmfgraph*}(10,10)
            \fmfleft{i}
            \fmfright{o}
            \fmf{plain}{i,v1}
            \fmf{plain}{o,v2}
            \fmffreeze
            \fmfforce{(0.1w,0.5h)}{v1}
            \fmfforce{(0.9w,0.5h)}{v2}
            \fmf{plain,left=1,tag=1}{v1,v2}
            \fmf{plain,right=1,tag=2}{v1,v2}
            \fmf{phantom,tag=3}{v1,v2}
            \fmf{phantom,tag=4}{w1,w2}
            \fmfforce{(0.1w,0.5h)}{w1}
            \fmfforce{(0.5w,0.3h)}{w2}
            \fmfposition
            \fmfipath{p[]}
            \fmfiset{p1}{vpath1(__v1,__v2)}
            \fmfiset{p2}{vpath2(__v2,__v1)}
            \fmfiset{p3}{vpath3(__v1,__v2)}
            \fmfiset{p4}{vpath4(__w1,__w2)}
            \fmfi{plain}{point 1*length(p1)/4 of p1 -- point 1*length(p3)/2 of p3}
            \fmfi{plain}{point 3*length(p1)/4 of p1 -- point 1*length(p3)/2 of p3}
            \fmfi{plain}{point 1*length(p3)/2 of p3 -- point 1*length(p2)/2 of p2}
            \fmfi{plain}{point 1*length(p1)/8 of p1 -- point 1*length(p4)/1 of p4}
        \end{fmfgraph*}
    }
}

\newcommand{\GraphTolopogyDfour}{
    \parbox[c][10mm][c]{11mm}{
        \centering
        \begin{fmfgraph*}(10,10)
            \fmfleft{i}
            \fmfright{o}
            \fmf{plain}{i,v1}
            \fmf{plain}{o,v2}
            \fmffreeze
            \fmfforce{(0.1w,0.5h)}{v1}
            \fmfforce{(0.9w,0.5h)}{v2}
            \fmf{plain,left=1,tag=1}{v1,v2}
            \fmf{plain,right=1,tag=2}{v1,v2}
            \fmf{plain,left=1,tag=3}{w1,w2}
            \fmf{plain,right=1,tag=4}{w1,w2}
            \fmfforce{(0.345w,0.5h)}{w1}
            \fmfforce{(0.655w,0.5h)}{w2}
            \fmfposition
            \fmfipath{p[]}
            \fmfiset{p1}{vpath1(__v1,__v2)}
            \fmfiset{p2}{vpath2(__v2,__v1)}
            \fmfiset{p3}{vpath3(__w2,__w1)}
            \fmfiset{p4}{vpath4(__w2,__w1)}
            \fmfi{plain}{point 1*length(p1)/4 of p1 .. point 1*length(p3)/4 of p3}
            \fmfi{plain}{point 3*length(p1)/4 of p1 .. point 3*length(p3)/4 of p3}
            \fmfi{plain}{point 1*length(p4)/2 of p4 .. point 1*length(p2)/2 of p2}
        \end{fmfgraph*}
    }
}

\newcommand{\GraphTolopogyCfour}{
    \parbox[c][10mm][c]{11mm}{
        \centering
        \begin{fmfgraph*}(10,10)
            \fmfleft{i}
            \fmfright{o}
            \fmf{plain}{i,v1}
            \fmf{plain}{o,v2}
            \fmffreeze
            \fmfforce{(0.1w,0.5h)}{v1}
            \fmfforce{(0.9w,0.5h)}{v2}
            \fmf{plain,left=1,tag=1}{v1,v2}
            \fmf{plain,right=1,tag=2}{v1,v2}
            \fmf{phantom,tag=3}{v1,v2}
            \fmfposition
            \fmfipath{p[]}
            \fmfiset{p1}{vpath1(__v1,__v2)}
            \fmfiset{p2}{vpath2(__v2,__v1)}
            \fmfiset{p3}{vpath3(__v1,__v2)}
            \fmfi{plain}{point 1*length(p1)/8 of p1 -- point 5*length(p2)/8 of p2}
            \fmfi{plain,rubout=3}{point 1*length(p3)/2 of p3 -- point 1*length(p2)/2 of p2}
            \fmfi{plain}{point 1*length(p1)/4 of p1 -- point 1*length(p3)/2 of p3}
            \fmfi{plain}{point 3*length(p1)/4 of p1 -- point 1*length(p3)/2 of p3}
        \end{fmfgraph*}
    }
}

\newcommand{\GraphTolopogyJfour}{
    \parbox[c][10mm][c]{11mm}{
        \centering
        \begin{fmfgraph*}(10,10)
            \fmfleft{i}
            \fmfright{o}
            \fmf{plain}{i,v1}
            \fmf{plain}{o,v2}
            \fmffreeze
            \fmfforce{(0.1w,0.5h)}{v1}
            \fmfforce{(0.9w,0.5h)}{v2}
            \fmf{plain,left=1,tag=1}{v1,v2}
            \fmf{plain,right=1,tag=2}{v1,v2}
            \fmfposition
            \fmfipath{p[]}
            \fmfiset{p1}{vpath1(__v1,__v2)}
            \fmfiset{p2}{vpath2(__v2,__v1)}
            \fmfi{plain,rubout=3}{point 1*length(p1)/4 of p1 -- point 2*length(p2)/4 of p2}
            \fmfi{plain,rubout=3}{point 2*length(p1)/4 of p1 -- point 1*length(p2)/4 of p2}
            \fmfi{plain,rubout=3}{point 2*length(p1)/3 of p1 -- point 2*length(p2)/3 of p2}
        \end{fmfgraph*}
    }
}

\newcommand{\GraphTolopogyBfour}{
    \parbox[c][10mm][c]{11mm}{
        \centering
        \begin{fmfgraph*}(10,10)
            \fmfleft{i}
            \fmfright{o}
            \fmf{plain}{i,v1}
            \fmf{plain}{o,v2}
            \fmffreeze
            \fmfforce{(0.1w,0.5h)}{v1}
            \fmfforce{(0.9w,0.5h)}{v2}
            \fmf{plain,left=1,tag=1}{v1,v2}
            \fmf{plain,right=1,tag=2}{v1,v2}
            \fmf{phantom,tag=3}{v1,v2}
            \fmfposition
            \fmfipath{p[]}
            \fmfiset{p1}{vpath1(__v1,__v2)}
            \fmfiset{p2}{vpath2(__v2,__v1)}
            \fmfiset{p3}{vpath3(__v2,__v1)}
            \fmfi{plain}{point 2*length(p1)/7 of p1 -- point 1.02*length(p3)/3 of p3}
            \fmfi{plain}{point 2*length(p1)/4 of p1 -- point 1.02*length(p3)/3 of p3}
            \fmfi{plain}{point 1.02*length(p3)/3 of p3 -- point 2*length(p2)/5 of p2}
            \fmfi{plain}{point 2*length(p1)/3 of p1 -- point 2*length(p2)/3 of p2}
        \end{fmfgraph*}
    }
}

\newcommand{\GraphTolopogyGfour}{
    \parbox[c][10mm][c]{11mm}{
        \centering
        \begin{fmfgraph*}(10,10)
            \fmfleft{i}
            \fmfright{o}
            \fmf{plain}{i,v1}
            \fmf{plain}{o,v2}
            \fmffreeze
            \fmfforce{(0.1w,0.5h)}{v1}
            \fmfforce{(0.9w,0.5h)}{v2}
            \fmf{plain,left=1,tag=1}{v1,v2}
            \fmf{plain,right=1,tag=2}{v1,v2}
            \fmf{plain}{w1,w2}
            \fmfforce{(0.33w,0.70h)}{w1}
            \fmfforce{(0.7w,0.3h)}{w2}
            \fmfposition
            \fmfipath{p[]}
            \fmfiset{p1}{vpath1(__v1,__v2)}
            \fmfiset{p2}{vpath2(__v2,__v1)}
            \fmfi{plain}{point 1*length(p1)/6 of p1 -- point 5*length(p1)/6 of p1}
            \fmfi{plain,rubout=3}{point 2*length(p1)/3 of p1 -- point 2*length(p2)/3 of p2}
        \end{fmfgraph*}
    }
}

\newcommand{\GraphTolopogyNfour}{
    \parbox[c][10mm][c]{11mm}{
        \centering
        \begin{fmfgraph*}(10,10)
            \fmfleft{i}
            \fmfright{o}
            \fmf{plain}{i,v1}
            \fmf{plain}{o,v2}
            \fmffreeze
            \fmfforce{(0.1w,0.5h)}{v1}
            \fmfforce{(0.9w,0.5h)}{v2}
            \fmf{plain,left=1,tag=1}{v1,v2}
            \fmf{plain,right=1,tag=2}{v1,v2}
            \fmfposition
            \fmfipath{p[]}
            \fmfiset{p1}{vpath1(__v1,__v2)}
            \fmfiset{p2}{vpath2(__v2,__v1)}
            \fmfi{plain,rubout=3}{point 1*length(p1)/4 of p1 -- point 2*length(p2)/4 of p2}
            \fmfi{plain,rubout=3}{point 2*length(p1)/4 of p1 -- point 3*length(p2)/4 of p2}
            \fmfi{plain,rubout=3}{point 3*length(p1)/4 of p1 -- point 1*length(p2)/4 of p2}
        \end{fmfgraph*}
    }
}

\newcommand{\GraphTolopogyMfour}{
    \parbox[c][10mm][c]{11mm}{
        \centering
        \begin{fmfgraph*}(10,10)
            \fmfleft{i}
            \fmfright{o}
            \fmf{plain}{i,v1}
            \fmf{plain}{o,v2}
            \fmffreeze
            \fmfforce{(0.1w,0.5h)}{v1}
            \fmfforce{(0.9w,0.5h)}{v2}
            \fmf{plain,left=1,tag=1}{v1,v2}
            \fmf{plain,right=1,tag=2}{v1,v2}
            \fmfposition
            \fmfipath{p[]}
            \fmfiset{p1}{vpath1(__v1,__v2)}
            \fmfiset{p2}{vpath2(__v2,__v1)}
            \fmfi{plain,rubout=3}{point 2*length(p1)/4 of p1 -- point 2*length(p2)/4 of p2}
            \fmfi{plain,rubout=3}{point 1*length(p1)/4 of p1 -- point 3*length(p2)/4 of p2}
            \fmfi{plain,rubout=3}{point 3*length(p1)/4 of p1 -- point 1*length(p2)/4 of p2}
        \end{fmfgraph*}
    }
}

\newcommand{\GraphTolopogyEfour}{
    \parbox[c][10mm][c]{11mm}{
        \centering
        \begin{fmfgraph*}(10,10)
            \fmfleft{i}
            \fmfright{o}
            \fmf{plain}{i,v1}
            \fmf{plain}{o,v2}
            \fmf{plain}{w1,w2}
            \fmffreeze
            \fmfforce{(0.1w,0.5h)}{v1}
            \fmfforce{(0.9w,0.5h)}{v2}
            \fmfforce{(0.5w,0.7h)}{w1}
            \fmfforce{(0.5w,0.3h)}{w2}
            \fmf{plain,left=1,tag=1}{v1,v2}
            \fmf{plain,right=1,tag=2}{v1,v2}
            \fmfposition
            \fmfipath{p[]}
            \fmfiset{p1}{vpath1(__v1,__v2)}
            \fmfiset{p2}{vpath2(__v2,__v1)}
            \fmfi{plain}{point 1*length(p1)/6 of p1 .. point 5*length(p1)/6 of p1}
            \fmfi{plain}{point 1*length(p2)/6 of p2 .. point 5*length(p2)/6 of p2}
        \end{fmfgraph*}
    }
}

\begin{document}

%%%%%%%%%%%%%%%%%%% COMS %%%%%%%%%%%%%%%%%%%
\newcommand{\com}[1]{\textcolor{magenta}{{#1}}} % com from Simon
\newcommand{\coms}[1]{\textcolor{red}{{#1}}} % com from Sofian

%%%%%%%%%%%% TEXT %%%%%%%%%%%%%%%%%%% 
\newcommand{\eg}{{\it e.g.}}
\newcommand{\ie}{{\it i.e.}}
\newcommand{\etal}{{\it et al.\ }}

%%%%%%%%%%%%% math basics %%%%%%%%%%%%%%%%%%%
\newcommand{\I}{\ensuremath{{\mathrm{i}}}}
\newcommand{\D}{\ensuremath{\mathrm{d}}}
\newcommand{\Tr}{\mbox{Tr}}
\newcommand{\Ord}{{\mathcal O}}
\newcommand{\eps}{\ensuremath{\varepsilon}}
\newcommand{\al}{\alpha}
\newcommand{\bra}{\langle }
\newcommand{\ket}{\rangle }
\newcommand{\abs}[1]{\left|#1\right|}
\def\ra{{\rightarrow}}
\newcommand{\fsl}[1]{{\centernot{#1}}} % feynman slash

%%%%%%%%%%%%% math special notations %%%%%%%%%%%%%%%%%%%
\newcommand{\EP}{\mathcal{E}} % EP
\newcommand{\xib}{\bar{\xi}} % modified gauge
\newcommand{\Lpt}{\tilde{L}} % modified momentum Log 
\newcommand{\KK}[1]{\mathcal{K}\!\left[\!\!\!#1\!\!\!\right]} % pole operator
\newcommand{\pE}{p_{{}_{\!E}}} % euclidean momentum
\newcommand{\kE}{k_{{}_{\!E}}} % euclidean momentum

%%%%%%%%%%%%% math ensembles %%%%%%%%%%%%%%%%%%%
\newcommand{\cE}{\ensuremath{\mathcal{E}}}
\newcommand{\cL}{\ensuremath{\mathcal{L}}}
\newcommand{\cC}{\ensuremath{\mathcal{C}}}
\newcommand{\cN}{\ensuremath{\mathcal{N}}}
\newcommand{\C}{\mathbb{C}}
\newcommand{\N}{\mathbb{N}}
\newcommand{\Q}{\mathbb{Q}}
\newcommand{\R}{\mathbb{R}}
\newcommand{\Z}{\mathbb{Z}}

%%%%%%%%%%%%%%%%%%% Math modes %%%%%%%%%%%%%%%%%%%
\def\be{\begin{equation}}
\def\ee{\end{equation}}
\def\ba#1\ea{\begin{align}#1\end{align}}
\def\bs{\begin{subequations}}
\def\es{\end{subequations}}
\def\nonum{\nonumber}

%%%%%%%%%%%%%%%%%%% Bigger parenthesis %%%%%%%%%%%%%%%%%%%
\makeatletter
\newcommand{\vast}{\bBigg@{4}}
\newcommand{\Vast}{\bBigg@{5}}
\makeatother

%%%%%%%%%%%%%%%%%%% OTHER %%%%%%%%%%%%%%%%%%%
\newcommand*\widefbox[1]{\fbox{\hspace{0.25cm}#1\hspace{0.25cm}}} % to box equations
\newcommand\titlemath[1]{\texorpdfstring{\(#1\)}{xx}} % for nice maths in titles 
\newcommand\phantomfrac{\phantom{$\cfrac{1}{1}$}\!\!\!} % just a fake fraction for alignement purposes
\newcommand\myclearpage{\checkoddpage\ifoddpage\else\newpage\mbox{}\fi}% clear pages of even page
\newcommand\mypart[1]{\myclearpage\part{\textsc{\textbf{#1}}}}

\def\lessspace{\!\!\!\!\!\!\!\!\!\!\!\!\!\!\!\!\!\!}

\newcommand{\hlmath}[1]{\colorbox{yellow}{$\displaystyle #1$}}
\newcommand{\HLmath}[1]{\colorbox{red}{$\displaystyle #1$}}
\newcommand{\HL}[1]{{\sethlcolor{red}\hl{#1}}}

\newcommand{\Plotetafit}{
\begin{tikzpicture}[scale=1]
\begin{axis}[
xmin = 0, xmax = 10,
ymin = 0.8, ymax = 1,
% xtick distance = 1,
% ytick distance = 5,
grid = both,
minor tick num = 1,
major grid style = {lightgray},
minor grid style = {lightgray!25},
width = 0.9\linewidth,
height = 0.6\linewidth,
legend cell align = {left}
]
\addplot[
domain = 0:10,
samples = 200,
smooth,
thick,
black,
] {0.834677 + 0.195333 * exp(-0.445534*x)};

\addplot[
domain = 0:10,
samples = 20,
smooth,
thick,
black,
dashed,
] {0.835};

\addplot[
smooth,
thick,
black,
only marks,
mark=square
] file[skip first] {figures/plot.dat};

\legend{$\eta_4=\eta_4^{\text{all-order}}+0.195 e^{-0.446 \ell} $, $\eta_4^{\text{all-order}}=0.835$}
\end{axis}
\end{tikzpicture}
}

\title{Four-loop elasticity renormalization of low-temperature flat polymerized membranes}
%\thanks{A footnote to the article title}%

\author{Simon Metayer$^{1,2}$}
%\email{simonmetayer@neuf.fr}
\affiliation{$^1$Institute of Nuclear and Particle Physics, Shanghai Jiao Tong University, Shanghai, China. \\
$^2$Laboratory of Theoretical and High Energy Physics, Sorbonne University, CNRS, Paris, France.}

\date{\today}

\begin{abstract}
We provide the complete four-loop perturbative renormalization of a low-temperature statistical mechanics model of flat polymerized membranes. Using a non-local effective flexural theory, which is based on transverse elastic fluctuations, we analytically determine the anomalous elasticity critical exponent $\eta$, which controls all scaling behaviors in the theory, at the four existing fixed points. The results are obtained as apparently convergent series, allowing for precise estimates without resummations. We independently verify and supplement the results of recent four-loop work [Pikelner 2022 EPL 138 17002] derived from a different model. Additionally, we find good agreement with non-perturbative theoretical approaches and experimental results on soft materials and graphene. 
\end{abstract}

%\keywords{Suggested keywords}%Use showkeys class option if keyword

\maketitle

%\tableofcontents

\section{Motivations}

This paper investigates the critical elastic properties of polymerized membranes, which are two-dimensional extensions of linear polymer chains, subject to small thermal fluctuations. Initially motivated by biophysics experiments that revealed the existence and stability of the flat phase in the cytoskeletons of red blood cells \cite{Schmidt:1993}, it has since attracted continued experimental interest, ranging from artificial soft materials such as amphiphilic films \cite{Gourier:1997} to recent studies on freely suspended graphene \cite{LopezPolin:2015,Nicholl:2017,LopezPolin:2022}.

A major challenge in this context is accurately determining the universal critical exponent $\eta$ at all infrared fixed points. This exponent governs the anomalous elasticity of the membrane, which controls all scaling behaviors in the flat phase, as reviewed in, \eg, \cite{Bowick:2001,Wiese:2020,Metayer:2024}. To achieve this, two equivalent theoretical models are used to renormalize thermal fluctuations: a two-field model that includes both in-plane (phonon) and out-of-plane (flexural) fields, and an effective flexural model, where in-plane modes are exactly integrated out. Since the early leading-order studies in small or large parameter expansions \cite{Aronovitz:1988,David:1988,Guitter:1988,Aronovitz:1989,Guitter:1989,Gornyi:2015,Saykin:2020} and the minimally self-consistent analysis \cite{Nelson:1987} \footnote{In \cite{Nelson:1987}, the authors analyze the simplest one-loop truncation of the self-consistent Schwinger-Dyson equations, where only the bending rigidity $\kappa$ is renormalized, while the couplings $\mu$ and $b$ remain scale-invariant.}, these models have continuously attracted attention for non-perturbative approaches like SCSA \cite{LeDoussal:1992,Gazit:2009,Zakharchenko:2010,Roldan:2011,LeDoussal:2018} and NPRG \cite{Kownacki:2009,Braghin:2010,Essafi:2011,Hasselmann:2011,Essafi:2014,Coquand:2016a,Coquand:2020a}, as well as Monte Carlo simulations \cite{Zhang:1993,Bowick:1996,Los:2009,Troster:2013} and recent interest in variations such as disordered extensions \cite{Coquand:2020b,Metayer:2022rdp} or chiral current effects \cite{Cardoso:2024vax}.

Advancing beyond leading order in perturbative approaches remained a challenge for three decades, until the two- and three-loop studies \cite{Coquand:2020a,Metayer:2021kxm} with computations conducted in both models. 
Surprisingly, the results for the shearless fixed point differed between the two models. 
According to \cite{LeDoussal:2018}, such a limit occurs, \eg, in nematic elastomer membranes \cite{Xing:2003,Xing:2003b,Stenull:2003,Stenull:2004,Xing:2005,Xing:2008,Xing:2008b}, whose physics is accurately captured only by the effective flexural model.
Recently, \cite{Pikelner:2021} achieved the four-loop renormalization for the two-field model only, thus missing the physical shearless fixed point. In this work, we independently derive the complete four-loop renormalization in the effective flexural model, providing a significant verification of \cite{Pikelner:2021} and capturing the physical shearless anomalous elasticity. 

%\newpage

\section{Effective flexural model}

We consider a $d$-dimensional membrane embedded in a larger $D$-dimensional Euclidean space. A massless field $h$ representing flexural phonon degrees of freedom describes the transverse elastic deformation (height) of the membrane. The action, in Fourier space, is \cite{Nelson:1987}
\ba
S& = \frac{1}{2} \int_{p} p^4 |h_\alpha(\vec p)|^2 +S_{\text{int}}\,,
\label{eq:S}
\ea
with $\int_p=\int\D^dp/(4\pi)^d$ and Greek indices $\alpha,\beta,...$ range from 1 to the codimension $d_c=D-d$. In \eqref{eq:S}, the effective non-local quartic interaction term arises from the exact integration over in-plane phonon degrees of freedom
\ba
S_{\text{int}}\!=\!\frac{1}{4}\int\hspace{-0.2cm}\int\hspace{-0.2cm}\int\hspace{-0.2cm}\int_{p_1,p_2,p_3,p_4}\hspace{-1.3cm} 
R_0(\vec p_5) h_\alpha(\vec p_1) h_\alpha(\vec p_2)h_\beta(\vec p_3) h_\beta(\vec p_4) \,, 
\label{eq:Sint}
\ea
with $\vec p_5=\vec p_1+\vec p_2=-\vec p_3-\vec p_4$ and the tensorial structure
\ba
\label{eq:MNTensors}
& R_0(\vec p_5) = \left(\mu M_{abcd}(\vec p_5) + b N_{abcd}(\vec p_5)\right)p_1^a p_2^b p_3^c p_4^d\,, \\
& M_{abcd}(\vec p)= \frac{1}{2} \left(P_{ac}P_{bd}+ P_{ad} P_{bc} \right) - N_{abcd}(\vec p), \nonum \\
& N_{abcd}(\vec p)= \frac{1}{d-1} P_{ab}P_{cd}, \quad P_{ab}= \delta_{ab}-\frac{p_a p_b}{p^2}\,, \nonum
\ea
where Latin indices $a,b,...$ range from $1$ to $d$. The theory involves two couplings: the shear modulus $\mu$ and $b$, which is a $d$-dimensional generalization of the Young modulus 
\be
b = \frac{\mu (d\lambda + 2\mu)}{\lambda + 2\mu},
\label{eq:bdef}
\ee
which is also proportional to the bulk modulus, with $\lambda$ and $\mu$ being the two Lamé coefficients from linear elasticity theory \cite{Landau:1959}. Although $b$ appears to depend on $\mu$, the tensors $M$ and $N$ are mutually orthogonal, satisfying $M_{abcd}N^{abcd}=0$. Following \cite{Coquand:2020a,Metayer:2021kxm}, we then assume that these two couplings renormalize independently and that both $b$ and $d_c$ are independent of $d$, which suffices to capture the model's physics \footnote{These assumptions are unnecessary in the two-field model, likely explaining the difference in anomalous elasticity at the shearless fixed point between the two models.}. Finally, note that the bending rigidity $\kappa$ and temperature $T$ have been absorbed into the field and couplings as $\mu\rightarrow 4\pi\mu\kappa^{2}/T$, $b\rightarrow 4\pi b\kappa^{2}/T$ and $h^2\rightarrow h^2T/\kappa$, all positive due to mechanical stability conditions.

\clearpage

\section{Renormalization setup}

We perform a perturbative expansion in the small couplings $\mu$ and $b$ using the dimensional regularization scheme around the upper critical dimension $d_{\text{uc}}$=4, with the following renormalization conventions
\be
h = Z^{1/2} h_r, \quad \mu = Z_\mu \mu_r M^{2\eps}, \quad b = Z_b b_r M^{2\eps}\,,
\ee
where $d=4-2\eps$ and the subscript $r$ denotes renormalized (running and finite) quantities. The multiplicative renormalization constants $Z$, $Z_\mu$ and $Z_b$ absorb all $\eps$-poles of the theory, employing the modified minimal subtraction scheme, with the renormalization scale $\overline{M}^{ 2} = 4\pi e^{-\gamma_E} M^2$. 

From the Dyson series, the dressed propagator \eqref{eq:S} gives
\be
\left\bra h_\alpha(\vec p)h_\beta(-\vec p)\right\ket=\frac{\delta_{\alpha\beta}}{p^4-\Sigma(\vec p)}\,,
\ee
and connects via the dressed quartic interaction
\ba
& \left\bra h_\alpha\!(\vec p_1\!)h_\beta\!(\vec p_2\!)h_\gamma\!(\vec p_3\!)h_\delta\!(\vec p_4\!)\right\ket=-2R(\vec p_5) \delta_{\alpha\beta}\delta_{\gamma\delta}\,, \nonum \\
& R(\vec p_5)=\left(\frac{\mu M_{abcd}(\vec p_5)}{1-\mu \Pi_M(\vec p_5)} + \frac{b N_{abcd}(\vec p_5)}{1-b \Pi_N(\vec p_5)}\right)p_1^a p_2^b p_3^c p_4^d\,,
\ea
where the self-energies $\Sigma$, $\Pi_M$ and $\Pi_N$ renormalize as
\ba
\label{eq:renormalizedselfenergies}
p^4-\Sigma_r(\vec p)& =(p^4-\Sigma(\vec p))Z \\
1-\mu_r\Pi_{Mr}(\vec p)& =-(1-\mu\Pi_{M}(\vec p))Z^{-2}Z_{\mu}^{-1}, \nonum \\
1-b_r\,\Pi_{Nr}(\vec p)\,& =-(1-b \Pi_{N}(\vec p))Z^{-2}Z_{b}^{-1}. \nonum 
\ea
Since the left-hand sides of the relations \eqref{eq:renormalizedselfenergies} are renormalized, they must be $\eps$-finite. This assertion defines our renormalization method, as it fully constraints the multiplicative renormalization constants $Z$, $Z_\mu$ and $Z_b$ as functions of the self-energies $\Sigma$, $\Pi_M$ and $\Pi_N$ at each loop order \footnote{Following the two- and three-loop studies \cite{Coquand:2020a,Metayer:2021kxm}, we find this renormalization technique more straightforward than conventional counter-term or BPHZ methods, as it does not require additional diagram computations.}. 

The renormalization group functions are defined as
\be
\eta = \frac{\D \log Z}{\D \log M} \,, \quad
\beta_\mu = \frac{\D \mu_r}{\D \log M}\,, \quad
\beta_b = \frac{\D b_r}{\D \log M}\,,
\label{ren-functions:def}
\ee
where the beta functions can be solved in the matrix form
\be
\hspace{-0.25cm}
\begin{bmatrix}
\beta_\mu \\
\beta_b \\
\end{bmatrix}
\!=\!-2\eps
\begin{bmatrix}
\mu_r \partial_{\mu_r}\log\mu_r Z_{\mu}
& \mu_r \partial_{b_r}\log \mu_r Z_{\mu} \\
b_r \partial_{\mu_r}\log b_r Z_{b}
& b_r \partial_{b_r}\log b_r Z_{b} \\
\end{bmatrix}
^{\!-1}\!\!\!\!
\begin{bmatrix}
\mu_r \\
b_r \\
\end{bmatrix}\,,
\ee
while the anomalous dimension of the field is given by
\be
\eta = \beta_\mu \partial_{\mu_r} \log Z +
\beta_b \partial_{b_r} \log Z .
\label{ren-functions:def+:gamma}
\ee
Solving $[\beta_\mu(\mu_r,b_r),\beta_b(\mu_r,b_r)]=0$ gives the coordinates of the fixed points $(\mu_n,b_n)$ for $n=1,2,3,4$ and the corresponding universal anomalous dimensions $\eta_n=\eta(\mu_n,b_n)$. 

We also recall that, according to scaling relations and Ward identities \cite{Aronovitz:1988,Guitter:1988,Aronovitz:1989,Guitter:1989}, the scale-invariant anomalous elasticity $\eta$ is the unique critical exponent that controls all other scaling behaviors in the theory, like
\be
\mu_r(p)\sim b_r(p) \sim p^{4-d-2\eta}\,, \quad \bra h_r(p)h_r(-p) \ket\sim p^{-4+\eta} ,
\ee
and can be computed from two-point massless diagrams with three-point interaction, using a Hubbard–Stratonovich transformation, see \cite{Metayer:2024}. 

\newpage

\section{Calculation details}

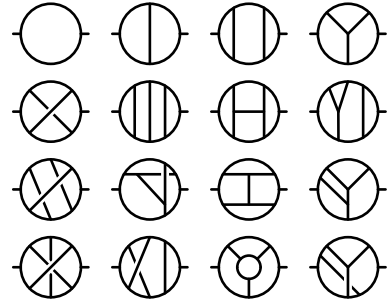
\begin{figure}[h!]
\centering
\begin{fmffile}{fmfmain}
\GraphTolopogyTone \GraphTolopogyTtwo \GraphTolopogyLthree \GraphTolopogyBthree \\
\GraphTolopogyNthree \GraphTolopogyLfour \GraphTolopogyHfour \GraphTolopogyBfour \\
\GraphTolopogyNfour \GraphTolopogyGfour \GraphTolopogyEfour \GraphTolopogyFfour \\
\GraphTolopogyMfour \GraphTolopogyJfour \GraphTolopogyDfour \GraphTolopogyCfour \\
\end{fmffile}
\caption{Propagator-type topologies up to four loops}
\label{fig:topologies}
\end{figure}

Up to four loops, there are $389$ Feynman diagrams, generated with \textsc{Qgraf} \cite{Nogueira:1993,Nogueira:2021wfp}, see details in table \ref{tab:diagsandintegrals}. Each diagram belongs to one of the $16$ maximum topologies of massless propagator types, illustrated in figure \ref{fig:topologies}. 

The numerator algebra is performed using custom \textsc{Mathematica} codes. Due to the intricate tensorial structure of the theory \eqref{eq:MNTensors}, each diagram results in a large momentum polynomial to be integrated, leading to the computation of over $57$ million multi-loop integrals. Using IBP reduction techniques \cite{Vasiliev:1981,Tkachov:1981,Chetyrkin:1981}, automated with both \textsc{LiteRed} \cite{Lee:2012cn,Lee:2013mka} and \textsc{Fire} \cite{Smirnov:2008iw,Smirnov:2013dia,Smirnov:2014hma,Smirnov:2019qkx}, 
all integrals are reduced to linear combinations of $39$ master integrals. This is the most time-consuming part, which is performed on a supercomputer. See details in table \ref{tab:diagsandintegrals}.
\begin{table}[h!]
\begin{ruledtabular}
\begin{tabular}{ccccc}
& 1-loop & 2-loop & 3-loop & 4-loop \\
$\Sigma$ diagrams & 1 & 3 & 18 & 155 \\
$\Pi_M$ diagrams & 1 & 2 & 12 & 91 \\
$\Pi_N$ diagrams & 1 & 2 & 12 & 91 \\
Topologies & 1 & 1 & 3 & 11 \\
Masters & 1 & 2 & 8 & 28 \\
Integrals & 15 & 796 & 191158 & 57245151
\end{tabular}
\caption{Number of diagrams and integrals computed.}
\label{tab:diagsandintegrals}
\end{ruledtabular}
\end{table}

Most of the masters can be partially or completely computed using massless integral techniques \cite{Kotikov:2018wxe}, while the remaining $16$ non-trivial integrals are provided, \eg, in \cite{Kazakov:1983,Kotikov:1996} at three loops and in \cite{Baikov:2010} at four loops. All masters are double-checked numerically using the sector decomposition tool \textsc{Fiesta} \cite{Smirnov:2009,Smirnov:2011,Smirnov:2014,Smirnov:2015mct,Smirnov:2021rhf} along with the Monte-Carlo integrator \textsc{Vegas} from the \textsc{Cuba} library \cite{Hahn:2004fe,Hahn:2014fua}. 

Typical results of the master integrals are asymptotic series expansions in $\eps$, including integer values of the Riemann zeta function $\zeta_n$, as illustrated by the expansion of the Euler Gamma function
\be
e^{\gamma_E \eps}\Gamma(\eps)=\frac{1}{\eps}+\frac{\zeta_2\eps}{2}-\frac{\zeta_3\eps^2}{3}+\frac{9\zeta_4\eps^3}{16}-\!\!\left(\!\!\!\!\frac{\zeta_2\zeta_3}{6}+\frac{\zeta_5}{5}\!\!\!\!\!\right)\eps^4+\Ord(\eps^5)\,,
\ee
up to a transcendental weight of $n=5$, which is the required order at four loops.

\clearpage

\section{Results}
\label{sec:results}

The complete set of renormalization group functions $\eta$, $\beta_\mu$, $\beta_b$, up to four loops and with full $d_c$ dependence, is provided in appendix \ref{sec:app:A}. From the beta functions $\beta_\mu$ and $\beta_b$, we extract the renormalization-group flow diagram, see figure \ref{fig:phasediag}, which reveals the expected total of 4 fixed points. The flow diagram remains qualitatively the same across all loop orders from one to four. 

Taking the anomalous dimension at these four infra-red fixed points, and displaying the result for simplicity in the physical case $d_c=1$, we obtain the following values for the anomalous stiffness critical exponents:

\newpage

\begin{figure}[h!]
\centering
\begin{picture}(50,50)
\put(0,0){\includegraphics[scale=0.5]{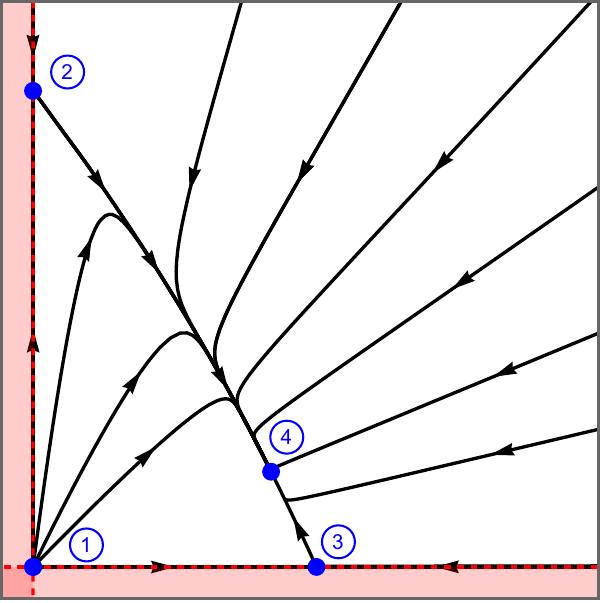}}
\put(47,4){$\mu_r$}
\put(4,47.5){$b_r$}
\end{picture}
\caption{RG-flow diagram in the ($\mu_r,b_r$) plane.}
\label{fig:phasediag}
\end{figure}

\vspace{-1cm}

\onecolumngrid
\begin{itemize}
\setlength{\itemindent}{-1.4em} 
\item Unstable Gaussian trivial fixed point $(\mu_1=b_1=0)$: 
%
%\vspace{-0.75cm}
\be
%\hspace{2cm} 
\eta_1 = 0 + \Ord(\eps^5).\\[-0.2cm]
\label{eq:eta1}
\ee
\item Semi-stable shearless fixed point $(\mu_2=0,b_2>0)$:
\be
\eta_2 = \frac{4 \eps}{5} - \frac{2 \eps^2}{375} + \frac{(119232 \zeta_3-120079) \eps^3}{2109375} + \frac{(-51994931+7803552 \zeta_3+26827200 \zeta_4+13512960 \zeta_5) \eps^4}{316406250}+ \Ord(\eps^5).\\[-0.0cm]
\label{eq:eta2}
\ee
\item Semi-stable infinitely compressible fixed point $(\mu_3>0,b_3=0)$:
\be
\eta_3 = \frac{20 \eps}{21} - \frac{94 \eps^2}{1323} - \frac{(312336 \zeta_3-9011) \eps^3}{5250987} - \frac{(14383003505+36705338304 \zeta_3+59031504 \zeta_4-56435313600 \zeta_5) \eps ^4}{661624362} + \Ord(\eps^5).\\[-0.2cm] 
\label{eq:eta3}
\ee
\item Fully stable fixed point $(\mu_4>0,b_4>0)$:
\be
\eta_4 = \frac{24 \eps}{25} - \frac{144 \eps^2\lessspace}{3125} - \frac{4(1286928 \zeta_3-568241) \eps^3\lessspace}{146484375} - \frac{4 (139409079893+355002697944 \zeta_3+723897000 \zeta_4-546469130880 \zeta_5) \eps ^4\lessspace}{54931640625} + \Ord(\eps^5). \\[-0.1cm]
\label{eq:eta4}
\ee
\end{itemize}
\twocolumngrid

\noindent These are exact order by order and numerically give
\ba
\eta_1 & = 0 + \Ord(\eps^5) \,, \label{eq:resultsnum}\\
\eta_2 & = 0.800 \eps - 0.005 \eps^2 + 0.011 \eps^3 + 0.001 \eps^4 + \Ord(\eps^5) \,, \nonum \\
\eta_3 & = 0.952 \eps - 0.071 \eps^2 - 0.070 \eps^3 - 0.075 \eps^4 + \Ord(\eps^5) \,, \nonum \\
\eta_4 & = 0.960 \eps - 0.046 \eps^2 - 0.027 \eps^3 -0.020 \eps^4 + \Ord(\eps^5) \,. \nonum 
\ea
The coefficients of \eqref{eq:resultsnum} are small and mostly decreasing, as if the series are convergent \footnote{All coefficients decrease, except for the three-loop contribution to the second fixed point and the four-loop contribution to the third fixed point, suggesting the onset of the expected asymptotic behavior of the series.}. This allows us to estimate the exact result by applying the optimal truncation rule \cite{Boyd:1999}, considering the physical case of a $d=2$ membrane embedded in a $D=3$ space ($\eps=1$), giving
\be
\eta_1 = 0 \,, \quad
\eta_2 = 0.807 \,, \quad
\eta_3 = 0.737 \,, \quad
\eta_4 = 0.867 \,.
\label{eq:rawresults}
\ee

For comparison, resummation estimates from Padé approximants yields $\eta_2^{[2/2]} = 0.807$, $\eta_3^{[2/2]} = 1.870$ and $\eta_4^{[2/2]} = 0.806$. On the one hand, the value at the second fixed point is very close (up to three digits) to the raw result \eqref{eq:rawresults}, indicating that despite the alternating signs, this series appears effectively convergent. On the other hand, the Padé approximant at the fourth fixed point is quite small, while at the third fixed point it falls outside the physical range $\eta\in[0,1]$. These approximations may then not be reliable, and we thus adhere to the raw values~\eqref{eq:rawresults}. 

\begin{figure}[h!]
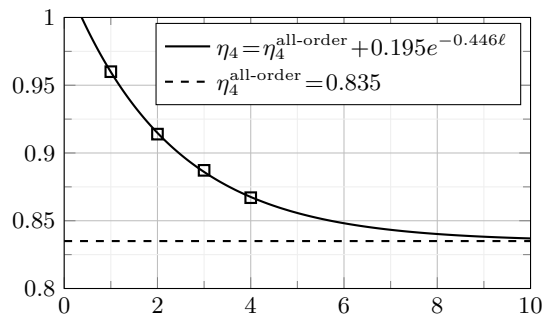

\centering
\Plotetafit
\caption{Exponential fit on $\eta_4$ from $1$ to $4$ loops ($\ell$).}
\label{fig:expfiteta}
\end{figure}

It is also interesting to observe that the successive loop correction coefficients to $\eta$ at the fully stable point appear to converge exponentially order by order, suggesting a fit, see figure \ref{fig:expfiteta}.

\noindent This leads to the conjecture of the all-order value 
\be
\eta_4^{\text{all-order}}=0.835 \,,
\label{eq:eta4best}
\ee
which is our best numerical estimate. A summary of these results is provided in Table \ref{tab:summary}.

\begin{table*}[ht]
\begin{ruledtabular}
\begin{tabular}{cccccccccc}
& 1-loop & 2-loop & 3-loop & 4-loop & Padé [2/2] & All-order & SCSA$^*$ & NPRG$^\dagger$ \\ 
$\eta_2$ & 0.800 & 0.795 & 0.806 & 0.807 & 0.807 & - & 0.821 & - \\
$\eta_3$ & 0.952 & 0.881 & 0.812 & 0.737 & - & - & - & - \\
$\eta_4$ & 0.960 & 0.914 & 0.887 & 0.867 & 0.806 & 0.835 & 0.821 & 0.849 \\
\end{tabular}
\end{ruledtabular}
\caption{Results from analytical works beyond leading order. $^*$\cite{LeDoussal:1992}, $^\dagger$\cite{Kownacki:2009}.}
%\vspace*{-0.5cm}
\label{tab:summary}
\end{table*}

\section{Sanity checks and discussion}

On the technical side, several non-trivial self-consistency checks are possible at such high loop order. First, a total of $113$ topological relations between diagrams are checked. Second, the leading pole polynomial structure of all diagrams in $\Pi_M$ and $\Pi_N$ should be identical, providing $106$ additional checks. Passing all these tests, together with the finiteness of the renormalization group functions and the expected vanishing of all terms proportional to $\zeta_2$, see appendix \ref{sec:app:A} and section \ref{sec:results}, are altogether an extremely strong support for our results.

Regarding exact comparisons with the literature, first and foremost, our four-loop contribution to $\eta_3$ and $\eta_4$ matches exactly with the recent results in \cite{Pikelner:2021}, which were computed in the independent two-field model, providing a significant and independent cross-check. Secondly, taking the large-$d_c$ limit of our results yields
\ba
& \eta_1 = 0 +\Ord(\eps^5/d_c^2)\,,\\
& \eta_2 = \frac{4}{d_c}\bigg(\!\!\!\eps+\frac{\eps^2}{6}-\frac{37\eps^3}{36} 
- \left(\!\!\frac{479}{216}-2\zeta_3 \!\!\!\!\!\right)\eps^4+\Ord(\eps^5)\!\!\bigg) + \Ord(1/d_c^2)\,, \nonum \\
& \eta_3 = \frac{20}{d_c}\bigg(\!\!\!\eps-\frac{37\eps^2\lessspace}{30}-\frac{31\eps^3\lessspace}{36}
-\left(\!\!\!\!\frac{769}{1080}-2\zeta_3\!\!\!\!\!\!\!\right)\eps^4+\Ord(\eps^5)\!\!\!\!\!\!\bigg) + \Ord(1/d_c^2)\,, \nonum \\
& \eta_4 =\frac{24}{d_c}\bigg(\!\!\!\eps-\eps^2-\frac{8\eps^3}{9}
-\left(\!\!\frac{26}{27}-2\zeta_3\!\!\!\!\right)\eps^4+\Ord(\eps^5)\!\!\bigg) + \Ord(1/d_c^2)\,, \nonum 
\ea
which aligns perfectly with the large-$d_c$ results $\forall d$ \cite{LeDoussal:1992,LeDoussal:2018} when re-expanded in $\eps$. Similarly, by neglecting all coupling corrections, we reduce the problem to only $72$ Feynman diagrams. At $d_c=1$, this analysis reveals
\ba
\eta_1 & = 0 + \Ord(\eps^5)\,, \\
\eta_2 & = \frac{4\eps}{5}-\frac{2\eps^2}{375} + \frac{697\eps^3}{28125} - \frac{(32400\zeta_3-36089)\eps^4}{843750} + \Ord(\eps^5)\,, \nonum \\
\eta_3 & = \frac{20\eps}{21} - \frac{206\eps^2}{3087} - \frac{25421\eps^3}{453789} + \frac{(2716560\zeta_3+3661109)\eps^4\lessspace}{133413966} + \Ord(\eps^5)\,, \nonum \\
\eta_4 & = \frac{24\eps}{25} - \frac{744\eps^2}{15625} -\frac{819016\eps^3}{29296875} - \frac{4(48262500\zeta_3-9441961)\eps^4\lessspace}{10986328125} + \Ord(\eps^5)\,, \nonum
\ea
again perfectly matching the findings of SCSA \cite{LeDoussal:1992,LeDoussal:2018} when re-expanded in $\eps$.

Additionally, our results show close agreement with non-perturbative literature. First, they all lie within the stability range $0<\eta<1$, which defines the physical flat phase bounded by the Gaussian ($\eta=0$) and minimally self-consistent ($\eta=1$) limits \cite{Nelson:1987}.
Secondly, we recall that the value of the shearless fixed point differs in the two-field model, with $\eta_2^{\text{two-field}}=0$ up to four loops \cite{Coquand:2020a,Metayer:2021kxm,Pikelner:2021}. We consider this trivial result as unphysical since such a shearless phase should be related, \eg, to the physics of nematic elastomers membranes \cite{Xing:2003,Xing:2003b,Stenull:2003,Stenull:2004,Xing:2005,Xing:2008,Xing:2008b}, where a positive anomalous elasticity has been observed. Indeed, our result \eqref{eq:rawresults} gives $\eta_{2}=0.867$, which is also compatible with SCSA, $\eta_{2}^{\text{SCSA}}=0.821$ \cite{LeDoussal:1992}.
Third, the infinitely compressible fixed point has no physical realization to our knowledge, although it is interesting for comparison purposes \cite{Metayer:2024} and has been shown to be conformal \cite{Guitter:1988,Mauri:2021}. 
Finally, for the fully-stable fixed point, our all-order numerical estimate \eqref{eq:eta4best} is $\eta_4^{\text{all-order}}=0.835$, which interestingly lies almost exactly between the estimates from NPRG, $\eta_4^{\text{NPRG}}=0.849$ \cite{Kownacki:2009} and SCSA, $\eta_{4}^{\text{SCSA}}=0.821$ \cite{LeDoussal:1992}, see table \ref{tab:summary}.
Experimentally, our results are consistent with soft-matter experiments $\eta_4^{\text{erythrocyte}}=0.7(2)$ \cite{Schmidt:1993} and $\eta_4^{\text{films}}=0.7(2)$ \cite{Gourier:1997}. We also find agreement with freely suspended graphene, supported by Monte Carlo simulations $\eta_4^{\text{MC}}\approx0.85$ \cite{Los:2009}, $\eta_4^{\text{MC}}=795(10)$ \cite{Troster:2013} and experiments $\eta_4^{\text{graphene}}\approx0.82$ \cite{LopezPolin:2015}.

\vspace{-0.25cm}
\section{Conclusion}

We computed the complete set of renormalization group functions of the effective flexural model for thermally fluctuating flat polymerized membranes, exactly up to four loops. We derived the renormalization-flow diagram and obtained the unique critical exponent of the theory, the anomalous elasticity $\eta$, for the four fixed points. These results are expressed as apparently convergent series, enabling precise numerical estimates and an all-order estimate at the fully stable fixed point. We find perfect agreement with previous results from large-$d_c$, SCSA, and partial four-loop results. Additionally, our results align well with all-order estimates from SCSA and NPRG, as well as recent Monte-Carlo simulations and graphene experiments.

\vspace{-0.25cm}
\begin{acknowledgments}
I am deeply grateful to Sofian Teber for our constant discussions, his crucial suggestions, for initiating the multiloop framework used in this work, and for his careful review of the manuscript. This work benefited from access to the HPC resources of the MeSU platform at Sorbonne University, which provided excellent technical support. I am grateful to Michela Petrini, head of LPTHE, for the funding approval of this work. I acknowledge Pierre Descombes for discussions on diagram~parametrizations. 
\end{acknowledgments}

\clearpage

\footnotesize

\bibliographystyle{apsrev4-1}

\bibliography{main}% Produces the bibliography via BibTeX.

\appendix

\onecolumngrid

%\clearpage

\vspace{-0.25cm}
\section{Renormalization-group functions up to four loops}
\label{sec:app:A}

\vspace{-0.55cm}

\footnotesize

\ba
\beta_\mu & =-2 \mu_r \eps +2 \mu_r \eta+\frac{d_c\mu_r^2}{6} + \frac{d_c\mu_r^2 \left(574\mu_r+107b_r\right)}{2^{4}3^{4}} +\frac{d_c\mu_r^2}{2^{9}3^{7}}\bigg[
\mu_r^2 (52 (409 d_c+7200)+321408 \zeta_3)
\\[-0.2cm] & 
+\mu_rb_r (224 (140 d_c+543)+88128 \zeta_3)
+b_r^2 (34987 d_c-176256 \zeta_3+113664)
\bigg] + \frac{d_c \mu_r^2}{2^{12}3^{10}} \bigg[ \mu_r^3 \big(8 (29169 d_c^2
\nonum \\[-0.1cm] & 
+4970049980 d_c+10779226092) -6912 \left(135 d_c^2-15314905 d_c-30857183\right) \zeta_3-186624 (65 d_c-1439) \zeta_4-622080 (258828 d_c+531313) \zeta_5\big)
\nonum \\ & 
+\mu_r^2 b_r \left(24 \left(224 d_c^2+685931199 d_c+657009061\right)+5184 (8443323 d_c+7107725) \zeta_3+46656 (317 d_c+3222) \zeta_4-933120 (71352 d_c+62111) \zeta_5\right)
\nonum \\ & 
+\mu_r b_r^2 \left(12 \left(48160 d_c^2+143406743 d_c+307944968\right)+20736 (222020 d_c+404643) \zeta_3+23328 (505 d_c-2892) \zeta_4-16796160 (416 d_c+795) \zeta_5\right) 
\nonum \\ & 
+b_r^3 \left(60235892 d_c+271405878-432 \left(6750 d_c^2-273983 d_c-1506511\right) \zeta_3+3059319 d_c^2-186624 (47 d_c+263) \zeta_4-311040 (582 d_c+3031)\zeta_5\right) 
\bigg] 
\nonum \\[-0.1cm] & 
+\Ord\Big(\!(\mu_r+b_r)^6\!\Big)
\nonum \\[-0.1cm] 
%%%%%%%%%%%%%%%%%%%%%%%%%%%%%%%%%%%%%%%%
\beta_b & = -2 b_r \eps +2 b_r \eta +\frac{5 d_c b_r^2}{12} + \frac{5 d_c b_r^2 \left(178\mu_r-91 b_r\right)}{2^{5}3^{4}} +\frac{d_cb_r^2}{2^{10}3^{7}} \bigg[
\mu_r^2 (4 (87893 d_c-248616)+1700352 \zeta_3)
\\[-0.2cm] & 
+\mu_rb_r (2240 (7 d_c-30)-725760 \zeta_3)
+b_r^2 (371495 d_c-228096 \zeta_3+614832)
\bigg] +\frac{d_c b_r^2}{2^{13}3^{10}} \bigg[\mu_r^3 \big(8 (34413 d_c^2
\nonum \\[-0.1cm] & 
+3552053866 d_c+6505946424)-3456 \left(1350 d_c^2-22155283 d_c-34097318\right) \zeta_3-186624 (316 d_c-7465) \zeta_4-1244160 (93111 d_c+150844) \zeta_5\big)
\nonum \\ &
+\mu_r^2 b_r \left(48 \left(1036 d_c^2+206496285 d_c+158637586\right)+5184 (5188449 d_c+3003346) \zeta_3+466560 (136 d_c+315) \zeta_4-1866240 (21915 d_c+13712) \zeta_5\right) 
\nonum \\ & 
+\mu_r b_r^2 \left(12 \left(33880 d_c^2+102216310 d_c+441948121\right)+2592 (1149311 d_c+3758781) \zeta_3+933120 (35 d_c-321) \zeta_4-16796160 (262 d_c+933) \zeta_5\right) 
\nonum \\ & 
+b_r^3 \left(23058315 d_c^2+31986562 d_c+123308526-432 \left(33750 d_c^2-32533 d_c-1226795\right) \zeta_3-466560 (31 d_c+148) \zeta_4-155520 (1446 d_c+5839) \zeta_5\right) 
\bigg] 
\nonum \\[-0.1cm] & 
+\Ord\Big(\!(\mu_r+b_r)^6\!\Big)
\nonum \\[-0.1cm] 
%%%%%%%%%%%%%%%%%%%%%%%%%%%%%%%%%%%%%%%%%%%%%%%%%%%%%
\eta & =\frac{5 (2 \mu_r+b_r)}{6} + \frac{-4\mu_r^2(111 d_c-20) +1160 \mu_r b_r+5 b_r^2 (15 d_c-212)}{2^{5}3^{4}} +\frac{1}{2^{9}3^{7}} \bigg[
\mu_r^3 \big(-8 \left(1395 d_c^2+188605 d_c-652398\right) 
\\[-0.1cm] & 
+ 10368 (96 d_c-509) \zeta_3\big)
+ \mu_r^2 b_r (12 (82681 d_c+108974)-15552 (50 d_c+57) \zeta_3)
+ \mu_r b_r^2 (6 (56445 d_c-221204)-108864 (5 d_c+2) \zeta_3)
\nonum \\[-0.1cm] & 
+ b_r^3 \left(-41625 d_c^2+180563 d_c+516252+5184 (9 d_c+16) \zeta_3\right)
\bigg]
+\frac{1}{2^{13}3^{10}} 
\bigg[ 
\mu_r^4 \big(
-16 \left(20763 d_c^3-445985328 d_c^2-15370535368 d_c-11680556284\right)
\nonum \\[-0.1cm] & 
+6912 \left(135 d_c^3+2722314 d_c^2+90983931 d_c+66318202\right) \zeta_3
+186624 \left(288 d_c^2+2003 d_c-20360\right) \zeta_4
-1244160 \left(23130 d_c^2+777029 d_c+569386\right) \zeta_5
\big)
\nonum \\ & 
+\mu_r^3 b_r \big(
16 \left(178219467 d_c^2+5548743275 d_c+8561040707\right)
+3456 \left(2196402 d_c^2+67864998 d_c+88562149\right) \zeta_3
-373248 \left(75 d_c^2+668 d_c+6800\right) \zeta_4
\nonum \\ & 
-622080 \left(18540 d_c^2+574178 d_c+778045\right) \zeta_5
\big) 
+ \mu_r^2 b_r^2\big(
24 \left(141866953 d_c^2+1890278050 d_c+1106116087\right)
\nonum \\ & 
+5184 \left(1737678 d_c^2+20174394 d_c+15225185\right) \zeta_3
-139968 \left(320 d_c^2+5433 d_c+3400\right) \zeta_4
-933120 \left(14634 d_c^2+176815 d_c+126437\right) \zeta_5
\big) 
\nonum \\ & 
+\mu_r b_r^3 \big(
20 \left(29547339 d_c^2+94268974 d_c+437788822\right)
+1728 \left(634335 d_c^2+4251180 d_c+9336377\right) \zeta_3
-2332800 \left(21 d_c^2+64 d_c+8\right) \zeta_4
\nonum \\ & 
-311040 \left(5760 d_c^2+30430 d_c+82109\right) \zeta_5
\big) 
+b_r^4 \big(
-8083125 d_c^3+26884554 d_c^2-71433074 d_c+126576784
\nonum \\ & 
+432 \left(16875 d_c^3-16896 d_c^2+627483 d_c+1500340\right) \zeta_3
+233280 \left(27 d_c^2+164 d_c+128\right) \zeta_4
+311040 \left(90 d_c^2-1991 d_c-4993\right)\zeta_5
\big)
\bigg]
+\Ord\Big(\!(\mu_r+b_r)^5\!\Big)
\nonum
\ea
\vspace{-1cm}
These results are available in computer readable files as ancillary files to the arXiv version of this letter.

\end{document}